*Research Article*

# A Novel Model of Cancer-Induced Peripheral Neuropathy and the Role of TRPA1 in Pain Transduction

**Ahmad Maqboul[1,2] and Bakheet Elsadek[2]**

[1]*Department of Anesthesiology and Operative Intensive Care Medicine, Campuses Mitte and Virchow-Klinikum, Charité–University of Medicine Berlin, Berlin, Germany*
[2]*Department of Biochemistry, College of Pharmacy, Al-Azhar University, Asyût, Egypt*

Correspondence should be addressed to Ahmad Maqboul; ahmad.maqboul@charite.de

*Background.* Models of cancer-induced neuropathy are designed by injecting cancer cells near the peripheral nerves. The interference of tissue-resident immune cells does not allow a direct contact with nerve fibres which affects the tumor microenvironment and the invasion process. *Methods.* Anaplastic tumor-1 (AT-1) cells were inoculated within the sciatic nerves (SNs) of male Copenhagen rats. Lumbar dorsal root ganglia (DRGs) and the SNs were collected on days 3, 7, 14, and 21. SN tissues were examined for morphological changes and DRG tissues for immunofluorescence, electrophoretic tendency, and mRNA quantification. Hypersensitivities to cold, mechanical, and thermal stimuli were determined. HC-030031, a selective TRPA1 antagonist, was used to treat cold allodynia. *Results.* Nociception thresholds were identified on day 6. Immunofluorescent micrographs showed overexpression of TRPA1 on days 7 and 14 and of CGRP on day 14 until day 21. Both TRPA1 and CGRP were coexpressed on the same cells. Immunoblots exhibited an increase in TRPA1 expression on day 14. TRPA1 mRNA underwent an increase on day 7 (normalized to 18S). Injection of HC-030031 transiently reversed the cold allodynia. *Conclusion.* A novel and a promising model of cancer-induced neuropathy was established, and the role of TRPA1 and CGRP in pain transduction was examined.

## 1. Introduction

Tumors spread through vascular, lymphatic, and/or neural routes. The latter is characteristic in cases of head and neck, pancreas, colon, rectum, prostate, biliary tract, and stomach cancers [1]. In these cases, 30% to 50% of patients suffer from moderate to severe pain [2]. The neuronal invasion varies from type to type. In pancreatic cancer, for example, the frequency is 90% or 100% [3]. The first model of neuropathic pain was developed in 1979 as a peripheral nerve injury [4]. Two years later, the chronic constriction injury "CCI" [5], which is still one of the most popular models for studying neuropathic pain in peripheral nerves, was developed.

Previously, cancer-induced neuropathy was modeled by injecting cancer cells in the proximity of the sciatic nerve (SN) [6, 7], promoting such models to study the invasion of malignant cells. However, the resident immune cells located in the surrounding tissue, such as the gluteal region, may interfere with evaluation of invasiveness and investigation of molecular mechanisms within the tumor microenvironment. To overcome the interference of these tissue-resident immune cells, which affects the onset of cancer pain and the regulation of the proteins involved in the invasion process, we introduce a novel cancer-induced neuropathic pain model. In this model, we directly inoculated the tumor cells within the neuronal sheath, which allowed a direct contact with the neuronal fibers from the time of injection.

To establish this model, we injected prostatic adenocarcinoma-derived cancer cells in the SN of Copenhagen rats. The cancer cells proliferated and generated spontaneous pain behavior and hypersensitivity, which later caused neuronal damage. We evaluated the tumor growth and measured the induced cold allodynia and mechanical and thermal hyperalgesia as indicators of pain development. Tissues were collected from the dorsal root ganglia (DRGs) to quantify the pain-recognition receptor, TRPA1, in parallel with the marker of tumor progress and the pain transducer, CGRP, and to study the underlying molecular mechanisms.

Transient receptor potential ankyrin 1 (TRPA1) plays a role in cold, mechanical, and thermal pain recognition. It is produced in a wide variety of tissues such as those in the central (CNS) [8] and peripheral (PNS) [9–12] nervous systems, eye [13], lung [14], gastrointestinal tract [15], genital systems [16], skin [17], cardiovascular system [18], urinary system [9, 10], and teeth [19, 20]. In some inflammatory conditions, such as streptozotocin-induced type I diabetes in rats [9] and cyclophosphamide-induced cystitis and hyperalgesia in mice bladders [10], TRPA1 is overexpressed by DRG neurons. In these tissues, the activation and the regulatory mechanisms of TRPA1 through local inflammation have been extensively investigated. However, TRPA1 regulation in cancer-induced peripheral neuropathy has not yet been studied.

Calcitonin gene-related peptide (CGRP), a 37-residue neuropeptide produced in sensory, motor, and autonomic neurons, is a potent vasodilator [21]. Vasodilation enhances the progression of cancer mainly through angiogenesis and lymph angiogenesis [22]. CGRP plays another role in transmitting signals of somatic, visceral, and neuropathic pain [23, 24]. Its production is correlated to the expression of TRPA1 in several inflammatory diseases such as colonic distension [25], carrageenan-induced chronic prostatitis and pelvic pain syndrome [26], and lipopolysaccharide-induced inflammation [27].

To confirm the nociceptive role of TRPA1 in our model of cold allodynia, we used HC-030031 to antagonize the TRPA1 channel. HC-030031 has a high selectivity for the TRPA1 receptor and is used for treating hyperalgesia in cyclophosphamide-produced [10], prostaglandin $E_2$-produced [11], and carrageenan-produced [12] pain models.

## 2. Methods

In our model, we used inbred male Copenhagen rats (Project Code: G 0314/13) for their major histocompatibility complex haplotype $RT1^{av1}$. This allows 100% growth of the transplantable tumor cells as previously reported [28]. These tumor cells, anaplastic tumor-1 (AT-1), were derived from the Dunning R3327 cell line and characterized by being androgen independent, nonmetastasizing, and having a very rapid growth rate [29]. The parental Dunning R3327 itself was derived from male Copenhagen rats (cf. thymic tumor which is more common in females) [30] and discovered as a spontaneous tumor in a 22-month-old male Copenhagen rat. It was first observed at a necropsy from the 54th brother × sister generation of its preceding cell line 2331 [31–33].

The study was approved by the State Office for Health and Social Affairs (LAGeSo, Berlin, Germany) and in adherence to the guidelines and the standards of Charité–University of Medicine Berlin.

### 2.1. Copenhagen Rats (COP/CrCrl).
Rats were obtained, through the Research Institutes for Experimental Medicine "FEM," Charité–University of Medicine Berlin, from Charles River Laboratories International, Inc., Cologne, Germany. They arrived with initial weights between 230 and 250 g and were housed at standard conditions of number per cage, food and water supply, and light exposure. A daily checkup was conducted to ensure normal life activities like moving, walking, playing, lying down, rising, and jumping. The degree of pain was scored during the whole research period.

### 2.2. Tumor Cell Injection and Tissue Collection.
We searched for different cancer cell lines that are capable of developing tumor in Copenhagen rats. Preliminary tests were performed for AR42J (ECACC, Salisbury, UK; Cat. #: 93100618) and AT-1 (ECACC; Cat. #: 94101449) cell lines.

Each rat was initially anesthetized with isoflurane and then maintained at 2% isoflurane in $O_2$ inhalation throughout the operation time. The right hind leg (the ipsilateral) was shaved and sterilized with alcohol and iodine. To expose the SN, an incision was made in the skin between the gluteus superficialis and the biceps femoris muscles.

The amphicrine AR42J cells were slowly inoculated $(0.5 \times 10^6)$ in the perineurium sheath of the SN of two successive groups of rats ($n = 5$ each) using a 25 μL Hamilton® GASTIGHT® syringe, 1702LT Series (Sigma-Aldrich Chemie GmbH) and watched for 40 days. Behavioral tests were measured on daily basis during the first two weeks and then every 5–7 days until the end of the observation period. We did not notice any tumor growth in SNs, and all animals recovered from surgery.

AT-1 cells $(1.0 \times 10^6)$ were then injected in another group of rats. During the first week, animals developed a spontaneous pain behavior and showed a varying degree of itching at the injection site. After being sacrificed, AT-1 cells showed a severe malignant growth outside the SN and in solid attachment to the muscles and the inner layers of the skin. Therefore, we reduced the number of injected AT-1 cells to $(0.5 \times 10^6)$. The growth rate of the tumor cells decreased with no invasion to the surrounding tissues. This number of cells was maintained for injection throughout the consecutive studies. Metamizole sodium was injected as a postsurgical analgesic and added to the drinking water for 3 days after. Tissues of lumbar L3–5 DRGs were collected and stored at −80°C as groups of sham at 3, 7, 14, and 21 days.

The anaplastic tumor AT-1 cells were cultivated in a medium of RPMI 1640, L-glutamine, dexamethasone, and 10% foetal bovine serum (FBS). Then, they were maintained at 37°C and 5% $CO_2$ atmosphere and counted under a light microscope using the Bright-Line™ Hemacytometer slide (Sigma-Aldrich Chemie GmbH, Steinheim, Germany). After suspending the cells in a pH-7.4 phosphate-buffered saline (PBS) as a vehicle, sham-operated rats were injected with this PBS and used as a control in all consecutive assays.

*2.3. Chemicals and Equipment.* Polyclonal rabbit anti-TRPA1 and monoclonal mouse anti-CGRP antibodies were purchased from Santa Cruz Biotechnology, Inc., Heidelberg, Germany (Cat. #: sc-66808), and Sigma-Aldrich Chemie GmbH (Product #: C7113), respectively; the corresponding secondary antibodies, Alexa Fluor® 594 donkey anti-rabbit IgG (Cat. #: A-21207) and Alexa Fluor 488 donkey anti-mouse IgG (Cat. #: A-21202), from Thermo Scientific™, IL, USA; the horseradish peroxidase- (HRP-) conjugated anti-rabbit from The Jackson Laboratory "Jax®," Sulzfeld, Germany (Cat. #: 111-035-144).

Other chemicals include isoflurane "Forene®" from AbbVie Deutschland GmbH & Co. KG, Ludwigshafen, Germany (Zul.-Nr.: 2594.00.00); paraformaldehyde from Sigma-Aldrich Chemie GmbH (P6148-500G); D(+)-Saccharose from Carl Roth GmbH + Co. KG, Karlsruhe, Germany (Art.-Nr.: 4621.1); Precision Plus Protein™ Kaleidoscope™ from Bio-Rad, München, Germany (Cat. #: 161-0375); bovine serum albumin (BSA) from Sigma-Aldrich, Co., MO, USA (PCode: 1001561418); 4′,6-diamidino-2-phenylindole (DAPI); ethanol "99%" and $\beta$-mercaptoethanol from Carl Roth GmbH + Co. KG, Karlsruhe, Germany; Novaminsulfon® Lichtenstein "500 mg metamizole sodium" drops from Winthrop Arzneimittel GmbH, Mülheim-Kärlich, Germany (Zul.-Nr. 6867259.00.00); 1,2,3,6-tetrahydro-1,3-dimethyl-N-[4-(1-methylethyl)phenyl]-2,6-dioxo-7H-purine-7-acetamide (HC-030031®) from Sigma-Aldrich Chemie GmbH (PCode: H4415); Pierce™ BCA Protein Assay Kit from Thermo Scientific; enhanced chemiluminescence "ECL" Kit from Amersham Pharmacia Biotech, NJ, USA; and Novaminsulfon-ratiopharm® "1 g/2 mL metamizole sodium ampoules" from ratiopharm GmbH, Ulm, Germany.

Equipment used in this study were Axiovert 25 inverted microscope (Carl Zeiss, Göttingen, Germany); Touch Test® Sensory Evaluators (North Coast Medical, Inc., CA, USA); Trans-Blot® Turbo™ Transfer System and nitrocellulose membranes (Bio-Rad, CA, USA); ChemiDoc™ MP Imaging System (Bio-Rad); CryoStar™ NX70 (Thermo Fisher Scientific™ Inc.); Mastercycler® Pro Thermal Cycler (Eppendorf AG, Hamburg, Germany); TaqMan™ 7500 Real-Time PCR System (Applied Biosystems, Thermo Scientific); UGO BASILE Plantar Test 37,370 (Ugo Basile® SRL, Varese, Italy); and Zeiss® LSM 510 laser scanning microscope (Carl Zeiss).

*2.4. Behavioral Assay*

*2.4.1. Cold Allodynia.* The acetone test was used to assess cold allodynia by exposing the animals ($n = 6$ for each group) to cold stimuli and counting their responses as the number of flicking or licking. Every single measurement was performed in a duplicate, with a 5-minute interval between each, and the average was calculated [34, 35]. The test was performed in cancer-injected and sham animals for baseline measurement. It was also performed at the 7th day after subcutaneously injecting the selective TRPA1 antagonist HC-030031 (at 15, 30, 60, 90, and 120 min). HC-030031 was dissolved in DMSO and then diluted in ethanol : water mixture (10 : 90 v/v) to give final concentrations of 25 and 50 mg/kg.

*2.4.2. Mechanical Hyperalgesia.* The Von Frey test was performed to measure hypersensitivity to mechanical stimuli by using Touch Test Sensory Evaluators (North Coast Medical, Inc.). The evaluators included filaments with logarithmic increments in force applied on the rats' hind paws ($n = 7$ for each group). The test was applied for the assessment of mechanical sensitivity in sham and cancer rats. The 50% g threshold was calculated using the up-and-down method with X = positive- and O = negative-responses according to Dixon's formula:

$$50\% \text{ g threshold} = \frac{\left(10^{[X_f + K\delta]}\right)}{10,000}, \quad (1)$$

where $X_f$ = the value of the last filament used (in log unit), $K$ = the tabular constant for OX series, and $\delta$ = the mean interval (in log unit; here = 0.229) [36, 37].

To calculate the interval ($\delta$), we calibrated the filaments by taking the mean of 10 measurements for each one. The tabular values ($K$) were provided as constants according to Dixon [38] and Chaplan et al. [37].

*2.4.3. Thermal Hyperalgesia.* The animal response to noxious heat stimuli was assessed by the Hargreaves method [39] using Ugo Basile Plantar Test. The withdrawal latency was measured in sham and cancer-inoculated animals ($n = 8$ for each group) as a mean of duplicate measurements (in seconds).

*2.5. Morphology and Light Microscope.* The shape of the ipsilateral hind paw in the cancer-injected rats was compared to the ipsilateral hind paw of the sham-operated rats and to the contralateral side of the same rat. After removal of the SNs, they were further checked at different time points for formation of malignant lumps and the gradual thickening and weight gain.

The invasion process and the morphological changes within the SNs were histologically examined by a light microscope. Tissues of the SNs from each group of sham and cancer animals were cut into sections (Section 2.7) and stained by hematoxylin and eosin. We visualized longitudinal sections of the SNs using an Axiovert 25 inverted microscope assembled with a CCD camera (AxioCam HRc).

*2.6. Gel Electrophoresis and Western Blotting.* TRPA1 was semiquantitatively assessed using Western blotting. The bilateral lumbar DRGs, L3–5, were collected from each group of animals and homogenized. DRG proteins were then extracted and measured using a Pierce BCA Protein Assay kit. They were added to a mixture of 2-mercaptoethanol and bromophenol blue and then separated on sodium dodecyl sulphate polyacrylamide gel (8%), using 7.5 $\mu$g protein per lane. The proteins were blotted on the nitrocellulose membrane using Trans-Blot Turbo Transfer System.

The blots were immersed in BSA for 90 min to block any nonspecific binding. Then, they were incubated overnight at 4°C with the anti-TRPA1 antibody (1 : 100) in BSA.

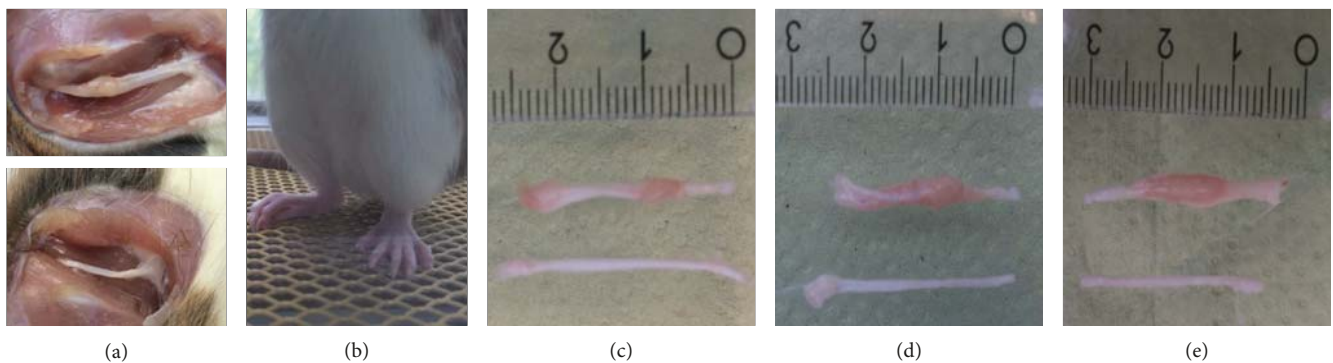

Figure 1: Morphological changes after AT-1 cell injection. (a) Sciatic nerves (SNs) are exposed in cancer (upper) and sham (lower) animals where the nerve thickness and the solid lump distinguish the cancerous one. (b) Signs of pain appear in the ipsilateral (right) and not in the contralateral (left) paw of the Copenhagen rat. Figures (c–e) show the gradual growth of the tumorous (upper) nerves in the cancer-injected animals at days 7 (c), 14 (d), and 21 (e). The nontumorous (lower) nerves are placed below for comparison.

We incubated the blots again for 120 min at room temperature with the secondary antibody. The secondary antibody used was horseradish peroxidase- (HRP-) conjugated anti-rabbit (1 : 40,000) in BSA. Then, the antigen-antibody complexes were visualized using the ECL kit. Images were captured by ChemiDoc MP Imaging System. To view the housekeeping protein band [40], we used the monoclonal anti-$\beta$-actin–peroxidase (1 : 25,000) in BSA. Fiji software was used to adjust the brightness and the contrast of images and to subtract the background and to invert the color [41]. The regions of interest (ROIs) were then selected for all the bands ($n = 11$ for each group) for which the integrated densities were measured. The integrated densities were then calculated as percentages to the corresponding control and compared to each other.

*2.7. Immunofluorescence Double-Staining.* To quantify TRPA1 and CGRP in DRGs, we used double-staining immunohistochemistry. Isoflurane was used to anaesthetize the rats, which were then transcardially perfused with 100 mL of a pH-7.4 PBS and then with 500 mL of 4% (w/v) paraformaldehyde in PBS. Lumbar L3–5 DRGs were collected in the same fixative and kept on the benchtop for 90 min. Then, they were washed with PBS and stored overnight at 4°C in 10% sucrose solution in PBS as a cryoprotectant. Next, they were embedded in an optimal-cutting-temperature (OCT) medium and stored at −20°C prior to cryostasis.

Tissues were cut into ultrathin sections ranging between 5 and 7 μm with CryoStar NX70. The sections were mounted onto gelatin-precoated slides and incubated overnight at room temperature with the primary anti-TRPA1 and anti-CGRP antibodies. Then, they were washed and incubated again with the corresponding secondary antibodies (Alexa Fluor 594 donkey anti-rabbit IgG and Alexa Fluor 488 donkey anti-mouse IgG, resp.). Nuclear DNA was labeled with DAPI according to the Chazotte protocol [42]. A laser scanning microscope, Zeiss LSM 510, was used to view the slides and to capture the photos ($n = 10$ for each group). These photos were used to count positive immunoreactive (+IR) cells relative to their total number (as percentage).

*2.8. Measurement of mRNA (qRT-PCR).* RNA was extracted from L3–5 DRGs of animals ($n = 9$ for each group) using an RNeasy kit (Qiagen, Hilden, Germany). Specific forward and reverse primers were used for cDNA synthesis. The quantification step was performed, by TaqMan 7500 Real-Time PCR System, using 18S ribosomal RNA as the reference normalization housekeeping gene [43].

*2.9. Statistical Data Analysis.* SigmaPlot 12.5 (Systat Software, Inc., CA, USA), GraphPad Prism (GraphPad Software, Inc., CA, USA), and Microsoft Excel 2016 (Microsoft Corp.) were used for doing statistics, graphs, and tabular data description. Data were expressed as mean ± standard error of mean (SEM). Results were tested for potential outliers. One-way analysis of variance (ANOVA) followed by Tukey's or Dunnett's post hoc test was used to make pairwise or versus control comparisons, respectively. The Kruskal–Wallis test was used to analyze variances on ranks if the normality test failed. Values of at least $P < 0.05$ were considered statistically significant.

## 3. Results

*3.1. Induction of Neuropathic Pain.* After inoculation of the tumor cells, we assessed the macroscopic and microscopic nerve characters and the pain behaviors. As expected, tumor growth took place only in the cancer-inoculated and not in the sham groups (Figure 1(a)). Moreover, the gradual proliferation of the tumor was only noticed in the ipsilateral and not in the contralateral hind paws (Figures 1(b)–1(e)).

Histology of the SNs was examined for evidence of tumor growth, nerve degeneration, and infiltration of immune cells. In sham animals, SN micrographs revealed normal fascicles with the characteristic curvy appearance and rod-like strands of nerve fibers. The nuclei of the Schwann cells and the fibrocytes were shown with no clear differentiation between them (Figure 2(a)).

In cancer-inoculated animals, the SNs were infiltrated by mononuclear cells. In addition, the gradual growth of the tumor from day 3 to day 21 was indicated by loss of integrity

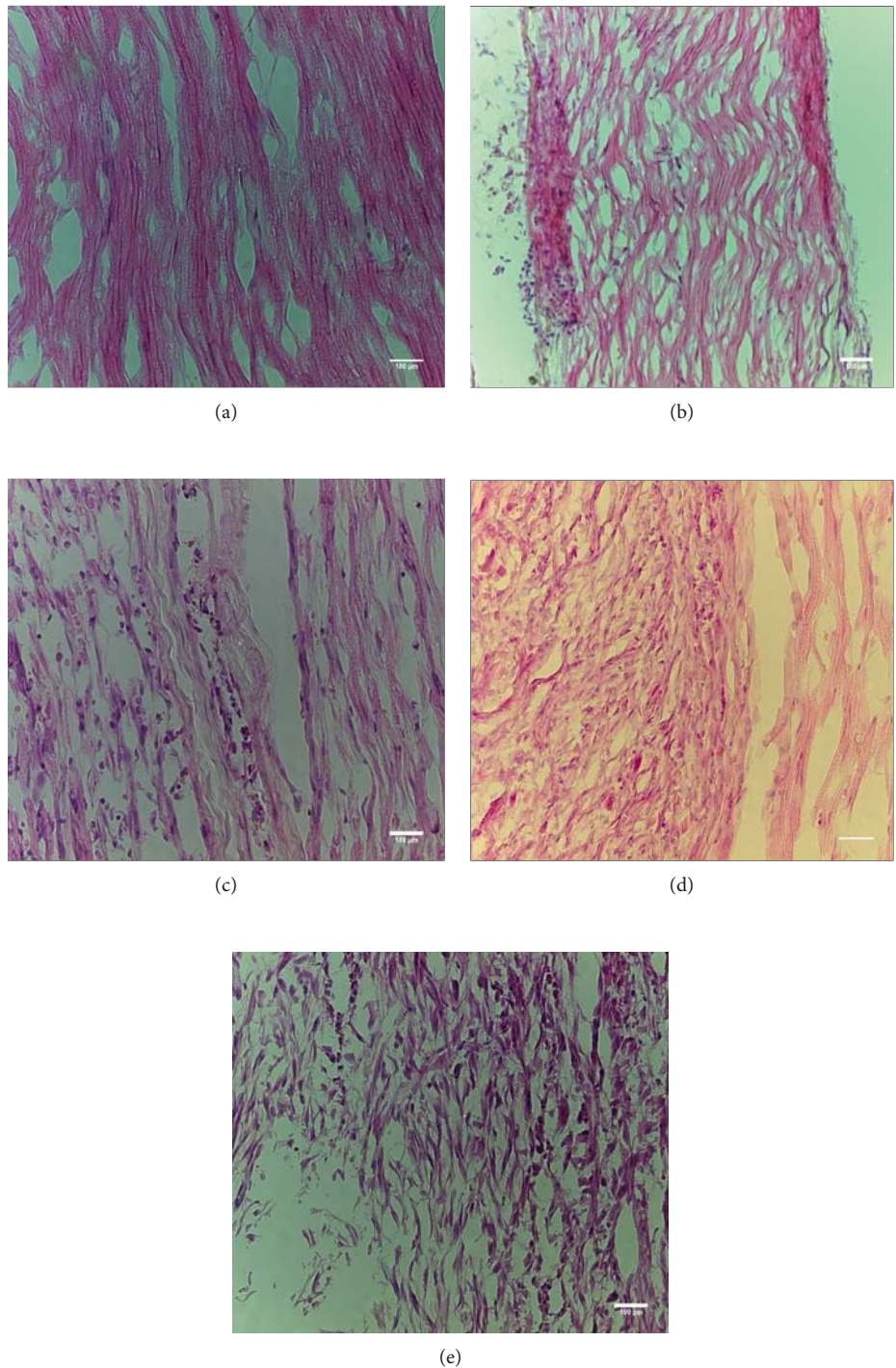

Figure 2: Light micrographs of the longitudinal sections of the sciatic nerve. (a) SNs of sham-operated animals show normal fascicles and a wavy pattern of the intact fibers which characterize normal nerves. (b–e) SNs of cancer-injected animals (at days 3, 7, 14, and 21, resp.) show a gradual invasion of cancer cells accompanied by a degradation of the nerve fibers. SNs are stained by hematoxylin and eosin (×20). Scale bars = 100 $\mu$m.

and degeneration of nerve fibers and the increasing number of nuclei over time (Figures 2(b)–2(e)).

The neuropathy was induced by injecting AT-1 cancer cells. Cold allodynia and mechanical and thermal hyperalgesia were estimated from day zero, "the baseline," to day fourteen in ipsilateral sides of both sham-operated and cancer-injected rats (Figure 3). The contralateral sides did not show any degree of neuropathic pain.

Cold allodynia was measured as the number of responses in sham and cancer animals ($n = 6$ for each group). In the sham group, the cold allodynia did not change during the observation period. In cancer animals, the hypersensitivity did not change until day 5 and then significantly increased at days 6 ($18 \pm 1.3$), 8 ($19 \pm 1.4$), 10 ($18 \pm 1.3$), 12 ($19 \pm 1.3$), and 14 ($20 \pm 1.7$) as compared to ($3.5 \pm 0.93$) for day zero (Figure 3(a)).

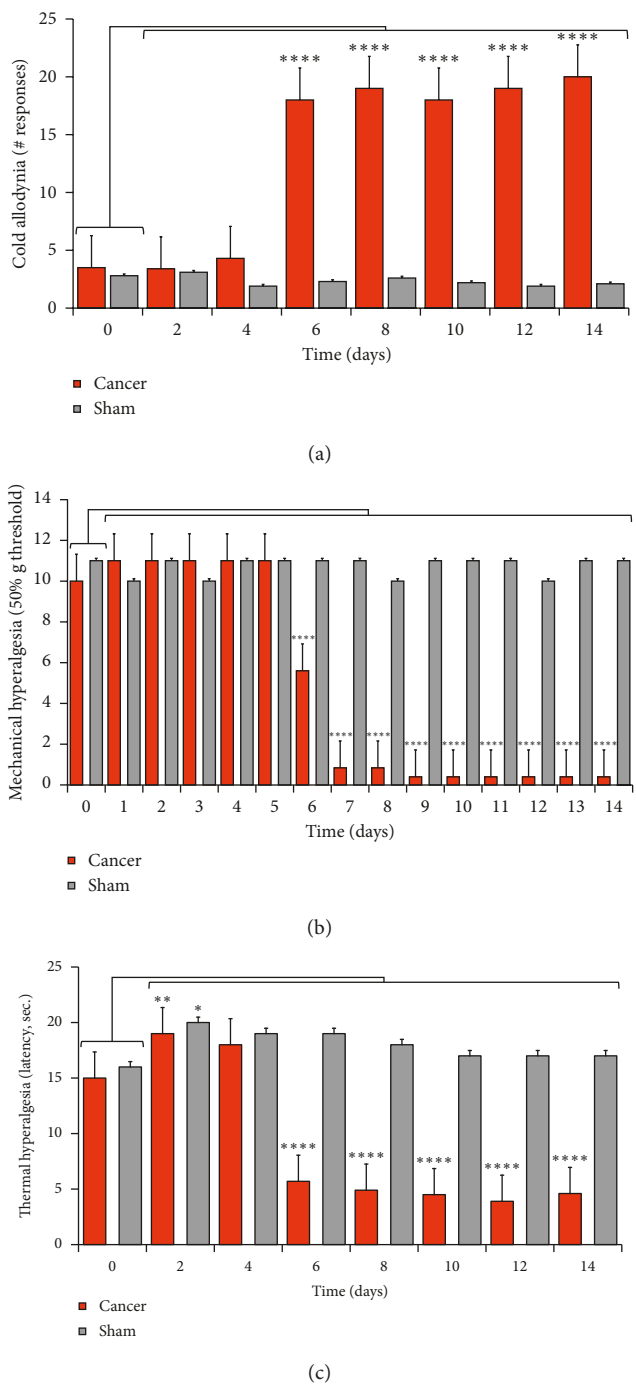

Figure 3: Cold allodynia and mechanical and thermal hyperalgesia. The number of responses to cold ($n = 6$), mechanical ($n = 7$), and thermal ($n = 8$) stimuli at the hind paw of the cancer animals is measured. (a) Substantial increase of cold allodynia from day 6 to day 14 ($P < 0.0001$) as indicated by an increased number of responses. (b) Significant increase of mechanical hyperalgesia from day 6 to day 14 ($P < 0.0001$) as indicated by decreased applied forces. (c) Thermal hyperalgesia significantly decreases at day 2 ($P < 0.01$) followed by an increase from day 6 to day 14 ($P < 0.0001$). Time points are all compared with control (day zero).

Mechanical pain measurements ($n = 7$ for each group) did not change in sham animals. In cancer animals, there was no change from day 1 to day 5. However, a significant increase in the degree of pain was observed from day 6 to day 14. The degree of pain was indicated by the decrease in applied force. The 50% g threshold dropped to ($5.6 \pm 1.9$) on day 6 and then to ($0.84 \pm 0.12$) during days 7 and 8 and reached a lowest value of ($0.40 \pm 0.0$) over days 9 to 14. All values were compared to day zero ($10 \pm 0.32$) (Figure 3(b)).

Hargreaves' test was performed ($n = 8$ for each group) for sham and cancer animals (Figure 3(c)). In sham animals, there was a decrease in the degree of nociception (i.e., an increased tolerance time) at day 2 ($20 \pm 1.0$) with ($P < 0.05$) as compared to the control ($16 \pm 0.7$). Then, it did not change till the end of the observation period. In cancer rats, the sensitivity to thermal stimuli decreased at day 2 ($19 \pm 0.86$) with ($P < 0.01$), did not change at day 4, and then increased at days 6 ($5.7 \pm 0.68$), 8 ($4.9 \pm 0.29$), 10 ($4.5 \pm 0.18$), 12 ($3.9 \pm 0.23$), and 14 ($4.6 \pm 0.26$) with ($P < 0.0001$) as compared to day zero ($15 \pm 0.77$).

*3.2. Upregulation of TRPA1 and CGRP in Afferent DRG Neurons.* Immunofluorescent images of DRG tissues were double-stained to show TRPA1 and CGRP expression as well as nuclei proliferation. The satellite cells in these images characterize the morphology of the peripheral neuroglia. Despite exhibiting similar upregulation, each had a different expression pattern.

On one hand, TRPA1 expression, as indicated by +IR cells, did not change till day 3. Then, it substantially increased until it reached its peak at day 7 ($P < 0.01$) and stayed significant until day 14 ($P < 0.05$). After day 14, it decreased at the same rate until it returned to its normal level at day 21. On the other hand, CGRP started with a slower rate of expression. It reached its maximum at day 14 and stayed at the same level till day 21 ($P < 0.001$). The nuclei stained with DAPI followed an increasing proliferation pattern similar to the CGRP one. It increased slowly from day 1 to day 7. Then, it dramatically increased at days 14 and 21 (Figure 4).

The upregulation of TRPA1 and CGRP and the proliferation of the nuclei as a response to the growing tumor were confirmed by the relative counting of the +IR cells (Figure 5). For TRPA1, the average of +IR cells was $42 \pm 2.8$ for the sham group, $55 \pm 3.4$ ($P < 0.01$) for the day 7 group, and $53 \pm 2.5$ ($P < 0.05$) for the day 14 group (Figure 5(a)).

The average of CGRP +IR cells was $35 \pm 2.2$ for the sham group, $52 \pm 3.2$ ($P < 0.001$) for the day 14 group, and $52 \pm 3.3$ ($P < 0.001$) for the day 21 group (Figure 5(b)). Nuclei counting revealed a significant increase at days 7 ($213 \pm 17$; $P < 0.05$), 14 ($225 \pm 11$; $P < 0.01$), and 21 ($244 \pm 12$; $P < 0.0001$) as compared to ($168 \pm 11$) of the sham group (Figure 5(c)).

The combined image of these three stains showed a coexpression of TRPA1 and CGRP on the same cells (Figure S1 in the supplementary file). In contrast to the sham group ($9.3 \pm 1.2$), the coexpression increased at day 3

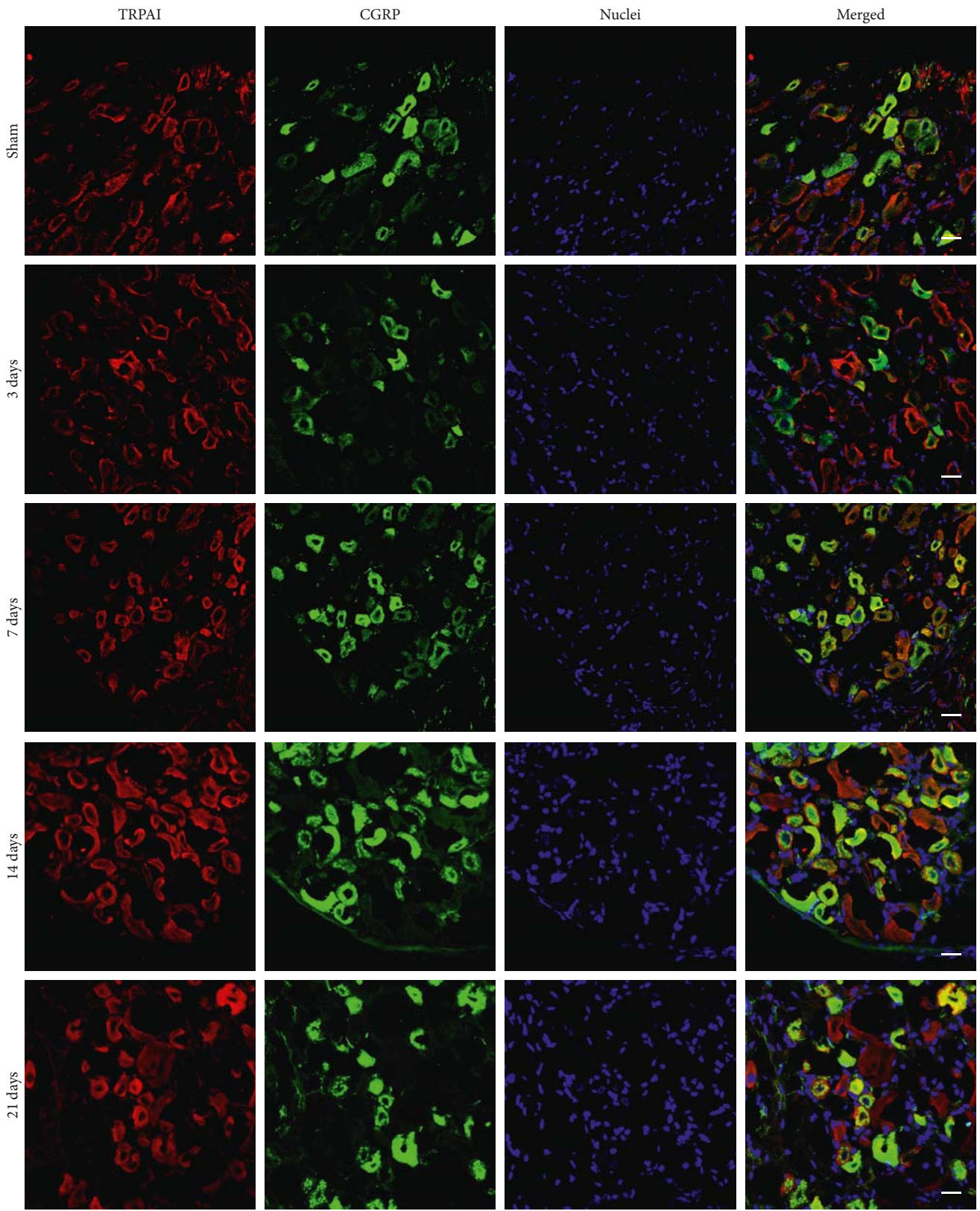

Figure 4: Double immunofluorescent micrographs of DRG neurons for control and 3-, 7-, 14-, and 21-day groups ($n = 10$). From left to right: TRPA1 is red, CGRP is green, nuclei (DAPI) is blue, and the merged is a combined image of the three. The yellow color indicates the coexpression of TRPA1 and CGRP on the same cells. Scale bars = 50 $\mu$m.

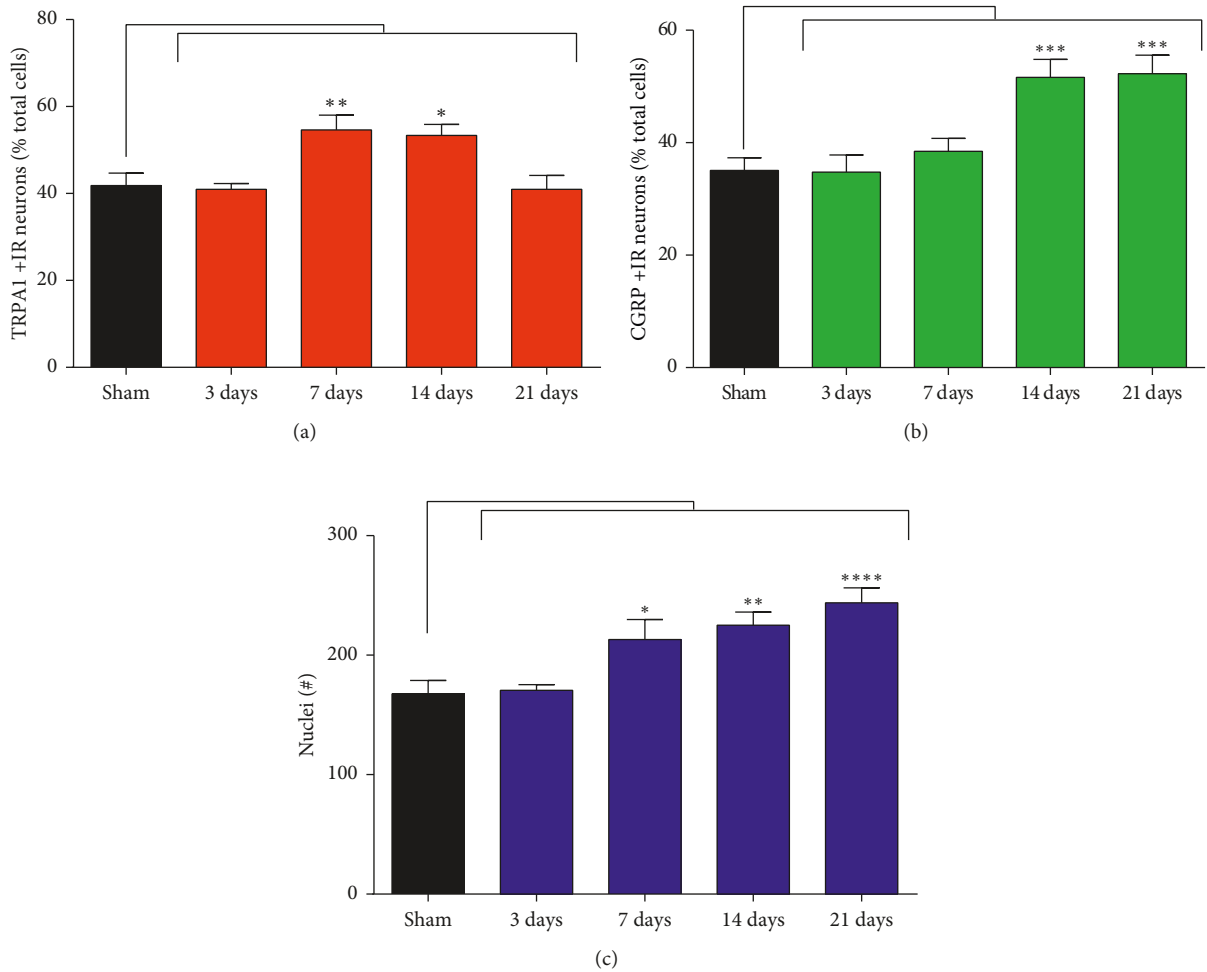

FIGURE 5: Quantification of TRPA1 and CGRP immunoreactive cells in DRGs as compared to the sham group. (a) TRPA1 shows a higher significance at day 7 ($P < 0.01$) and then a less at day 14 ($P < 0.05$). (b) CGRP shows a high significance at days 14 and 21 ($P < 0.001$). The numbers of +IR cells is expressed as percentage of total cells. (c) DAPI-stained nuclei are increased at days 7 ($P < 0.05$), 14 ($P < 0.01$), and 21 ($P < 0.0001$).

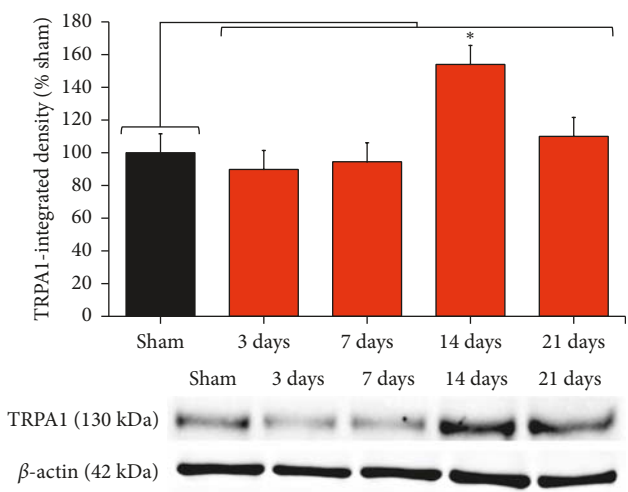

FIGURE 6: Immunoblots and quantitative densitometry of TRPA1. The bands show TRPA1 (130 kDa) and the internal standard, $\beta$-actin (42 kDa), blotted. The mean of the integrated density of each group of samples ($n = 11$) was measured and expressed as a percentage relative to that of the sham group.

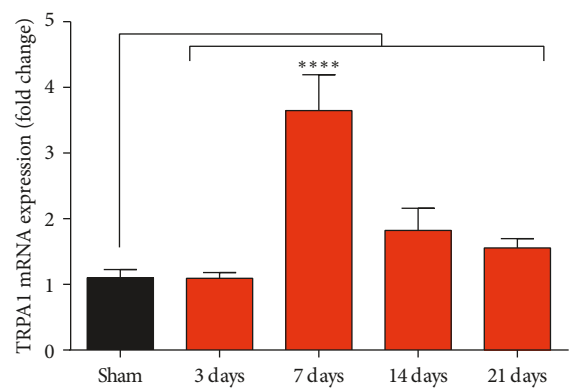

FIGURE 7: Quantification of TRPA1 mRNA in control and 3-, 7-, 14-, and 21-day groups ($n = 9$). Results are calculated and normalized to 18S rRNA as a preferred housekeeping gene.

($19 \pm 1.9$; $P < 0.05$) and became more significant at day 7 ($38 \pm 4.9$; $P < 0.0001$). It then decreased at day 14 ($28 \pm 3.0$; $P < 0.0001$) and day 21 ($23 \pm 2.4$; $P < 0.01$).

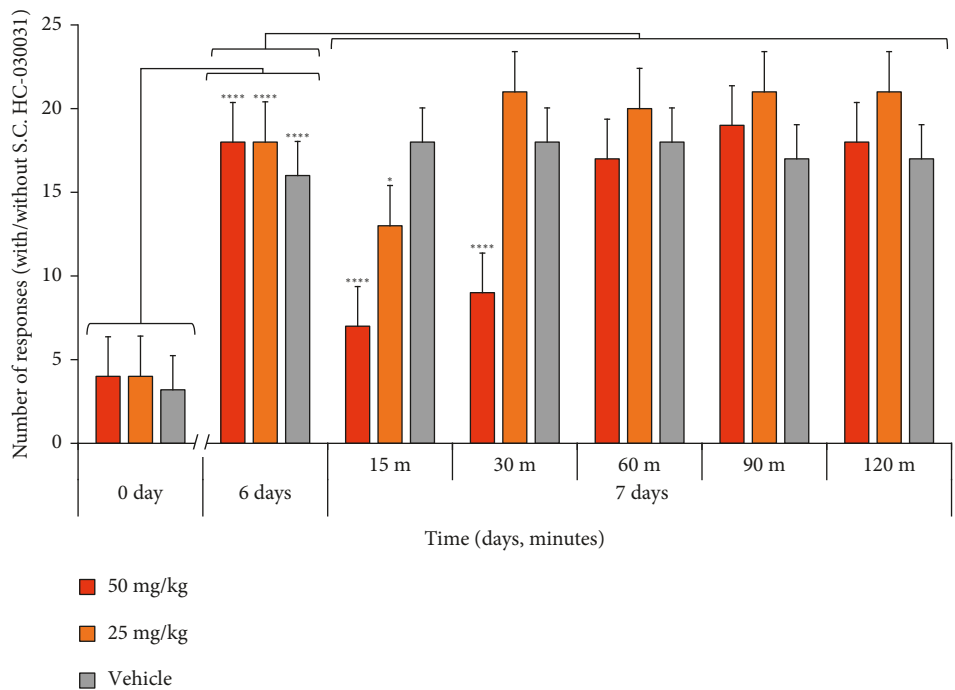

Figure 8: Cancer-induced neuropathy and its reversal using the selective TRPA1 antagonist HC-030031: the neuropathic pain (as indicated by cold allodynia) and its reversal in cancer animals injected with 50 and 25 mg/kg and vehicle ($n = 6$ for each group). The HC-030031 dose of 25 mg/kg restored the cold allodynic pain at 15 minutes ($P < 0.05$), while the 50 mg/kg dose showed a high significance ($P < 0.0001$) at 15 and 30 minutes.

The blots showed a trend of TRPA1 expression similar to that in the immunofluorescence images (see Figure S2 and S3 in the supplementary file). They did not change at days 3 and 7 and then significantly increased at day 14 ($154 \pm 25$; $P < 0.05$) and reversed at day 21 ($110 \pm 8.34$) as compared to sham ($100 \pm 0.0$) (Figure 6).

The real-time polymerase chain reaction results showed the same pattern as immunohistochemistry images. The expression values of TRPA1 mRNA increased at day 7 ($3.6 \pm 0.55$; $P < 0.0001$) and then became lower at day 14 and were restored at day 21 (Figure 7).

### 3.3. Transient Reversal of Cold Allodynia.
To study the effect of the TRPA1 selective antagonist HC-030031, we injected the animals with two doses of HC-030031 (25 and 50 mg/kg). The number of responses was measured at 15, 30, 60, 90, and 120 minutes and compared to the maximum of hypersensitivity reached previously at day 6 (Figure 3(a)).

For the 25 mg/kg dose, the number of responses decreased to ($13 \pm 0.9$; $P < 0.05$) at minute 15. The pain relapsed quickly at minute 30, and the number of responses increased again to ($21 \pm 1.0$) and plateaued to the end of measurement.

The 50 mg/kg dose had a longer lifetime as the number of responses significantly decreased to ($7.0 \pm 0.8$; $P < 0.0001$) at minute 15 and stayed significant till minute 30 ($9.0 \pm 0.5$; $P < 0.0001$). The pain relapsed again, and the number of responses increased to ($17 \pm 1.0$) at minute 60, which was lower than that of the 25 mg/kg dose at the same time point (Figure 8).

## 4. Discussion

Most of the established cancer-induced pain models are bone cancers. The pain behaviors in these models were reported at three [44] or four [45, 46] weeks after injection time in mice and around one week in rats [47]. To our knowledge, no cancer-induced neuropathic pain models were designed by direct inoculation of cancer cells within a peripheral nerve. In the current study, we introduce a novel model of cancer-induced neuropathy developed by direct injection of tumor cells into the neuronal sheath of the sciatic nerve.

In a previous study, cancer cells were injected in the close vicinity of the sciatic nerve in mice. Mechanical and thermal hyperalgesia was reported at days 10 to 14. This hyperalgesia was correlated with a mild degeneration of myelin at day 10 followed by a progressive degeneration until day 14 [6]. A similar model reported a significant mechanical hyperalgesia at day 14 and day 17. In this study, sensation was lost at day 21 [7].

In our model, tumor growth was morphologically verified by the constant increase in size and thickness of the anaplastic tumor-1 (AT-1) cell-injected sciatic nerves and the continuous formation of malignant lumps. The invasiveness of the cells was minimal as compared to other models, and hence, the duration of pain was longer [6, 7]. Microscopic examination confirmed infiltration of mononuclear cells throughout the fascicular mass as well as degeneration of nerve fibers. Evidence of neuropathic pain was shown by the progressive cold allodynia and mechanical and thermal hyperalgesia as compared to sham-operated animals.

Upon the development of such a local nonmetastasizing malignancy, the afferent DRG sensory neurons showed major changes on the molecular level. On one hand, the overexpression of the transient receptor potential ankyrin 1 (TRPA1) in the DRG itself mediates molecular responses to cancer development [19, 20] and transduces cold stimuli to the nociceptor sensory neurons. We observed that this overexpression is more significant upon inoculating tumor cells within the neuronal sheath. On the other hand, the number of calcitonin gene-related peptide (CGRP) +IR cells shows a nearly similar attitude of increased expression in response to the distant tumor growth. This is consistent with previous studies confirming its release in cancer and other chronic painful inflammatory conditions [48–50]. The delayed release of CGRP (at day 14) than TRPA1 (at day 7) suggests the dependence of CGRP production on TRPA1.

The concurrent expression of CGRP by TRPA1 +IR neurons increases with continuous tumor growth and neuronal damage. This implies a mutual synergistic function of pain transduction. Although Barabas and colleagues reported previously that the majority of TRPA1 +IR cells were CGRP-negative [51], our results are in agreement with other studies showing a high coexistence of TRPA1 and CGRP +IR neurons on the same cells [52–54].

The fluorescence images of TRPA1 +IR cells comply with a significant increase in TRPA1 mRNA at the same time point as well as with a similar trend of quantitative protein analysis using Western blot.

The noticeable increase in the number of nuclei in fluorescence images is attributed to the recruitment of immune cells to the nociceptive neurons. Most of the stained nuclei appear to be nonneuronal because of the nonspecific nature of the counterstain DAPI, which labels TRPA1 and CGRP +IR cells as well as nonneuronal cells [55, 56].

With the continuous growth of cancer cells, pressure is gradually exerted on the injection site of the sciatic nerve. This pressure leads to an increase in the hypersensitivity to cold stimuli as mentioned above. After the subcutaneous injection of the selective TRPA1 antagonist HC-030031, the cold allodynia is substantially suppressed, and the effect of cancer-induced neuropathy is transiently reversed in a dose-dependent manner. Transiently restoring the pain was formerly achieved using the same antagonist in other types of hypersensitivity [10–12, 19].

## 5. Conclusion

We successfully established a novel and a promising animal model of perineural cancer cell invasion in Copenhagen rats. This model was used for assessing cancer-induced neuropathic pain (i.e., cold allodynia and mechanical and thermal hyperalgesia) dependent on the degree of invasion. We also investigated the molecular mechanisms underlying the perineural invasion process and identified the role of the pain recognition receptor TRPA1 and the pain transmitter and vasodilator CGRP and highlighted their upregulation and coexpression in the dorsal root ganglionic neurons. Finally, we evaluated a therapeutic strategy for the relief of this perineural cancer pain by blocking the TRPA1 channel.

## Conflicts of Interest

The authors declare no conflicts of interest.

## Authors' Contributions

Ahmad Maqboul designed the study, performed the experiments, analyzed the results, and wrote the manuscript. Bakheet Elsadek contributed in the study design and writing and reviewing the manuscript.

## Acknowledgments

Thanks are due to Charité–University of Medicine Berlin for providing the research laboratories at the Department of Anesthesiology and Operative Intensive Care Medicine. Al-Azhar University is also acknowledged for supporting this study. This work is financially supported by the Egyptian Ministry of Higher Education (Cultural Affairs and Missions Sector) for two years and the German Foundation of Professor K. H. René Koczorek (Grant no.: IA89838780) for one year.

## Supplementary Materials

Figure S1: Quantification of TRPA1 and CGRP coexpression. Figure S2: Anti-TRPA1 antibody (Santa Cruz, sc-66808), 130 kDa. Figure S3: Anti-$\beta$-actin antibody (Sigma, A3854), 42 kDa. (*Supplementary Materials*)